\documentclass[12pt]{article}

\usepackage{color}

\usepackage{epstopdf}

\usepackage{graphicx}

\begin{document}

\begin{center}
{\bf Invariant Solutions of the Two-Dimensional Shallow Water Equations
with a Particular Class of Bottoms}\footnote{Submitted to AIP Conference Proceedings 2164, 050003 (2019)}

%
\vspace*{5mm}

{S.V. Meleshko$^\dag$\footnote{Corresponding author: sergey@math.sut.ac.th} and N.F. Samatova$^\ddag$}

{$^\dag$ School of Mathematics, Institute of Science, Suranaree University of Technology,
Nakhon Ratchasima}

{$^\ddag$ Department of Computer Science, College of Engineering,\\
 NC
State University, US}



\end{center}

{\bf Abstract}
The two-dimensional shallow water equations with a particular bottom and the Coriolis's force $f=f_{0}+\Omega y$ are studied in this paper. The main goal of the paper is to describe all invariant solutions for which the reduced system is a system of ordinary differential equations. For solving the systems of ordinary differential equations we use the sixth-order Runge-Kutta method.

\section{INTRODUCTION}

The two-dimensional shallow water equations with a Coriolis's force
$f=f_{0}+\Omega y$ have the form

\begin{equation}
\begin{array}{c}
h_{t}+uh_{x}+vh_{y}+h(u_{x}+v_{y})=0,\\
u_{t}+uu_{x}+u_{y}v-fv+2h_{x}=B_{x},\\
v_{t}+uv_{x}+vv_{y}+fu+2h_{y}=B_{y},
\end{array}\label{eq:June24.1}
\end{equation}
where $h$ is the deepness of the layer of fluid, $(u,v)$ is the
velocity, $t$ is time, $(x,y)$ are Eulerian coordinates, $-B(x,y)$
is the bottom, $f$ is the Coriolis's force. Group properties of the
one-dimensional equations (\ref{eq:June24.1}) with $f=0$ were studied
in \cite{bk:AksenovDruzhkov[2016]}. In \cite{bk:MeleshkoSamatova2019},
the authors studied group classification of the two-dimensional shallow
water equations in mass Lagrangian coordinates $(t,\xi,\eta)$ which
are introduced by the relations
\[
\dot{x}=u(t,x,y),\,\,\,\dot{y}=v(t,x,y),\,\,\,h=J^{-1},
\]
where $x=x(t,\xi,\eta)$, $y=y(t,\xi,\eta)$, $J=x_{\xi}y_{\eta}-x_{\eta}y_{\xi}$,
subindexes mean derivatives with respect to $\xi$ and $\eta$, the
sign $\dot{}$ means the derivative with respect to $t$. The advantage
of the Lagrangian coordinates is that the differential equations corresponding
to (\ref{eq:June24.1}) have a variational structure with the Lagrangian
\[
{\cal L}=\frac{1}{2}(\dot{x}^{2}+\dot{y}^{2})-(J^{-1}-B)+x(f_{0}+\Omega y)\dot{y.}
\]
The variational structure allowed the authors of \cite{bk:BilaMansfieldClarkson2006,bk:KaptsovMeleshko2019}
to apply Noether's theorem \cite{bk:Noether[1918]}
for deriving conservations laws. In \cite{bk:BilaMansfieldClarkson2006},
the group analysis method was applied to the two-dimensional shallow
water equations in mass Lagrangian coordinates with the flat bottom
$B=0$ and $f_{0}^{2}+\Omega^{2}\neq0$. For $f=0$ and $B=0$
equations (\ref{eq:June24.1}) coincide with the equations describing a two-dimensional
isentropic flow of a polytropic gas with the polytropic exponent $\gamma=2$.
Group properties of the two dimensional gas dynamics equations of
a polytropic gas in mass Lagrangian coordinates were analyzed in \cite{bk:KaptsovMeleshko2019}.

It should be also noted the paper \cite{bk:KaptsovMeleshko2019_2}, where flows of one-dimensional
continuum in Lagrangian coordinates are considered. Equations describing
these flows are reduced to a single Euler-Lagrange equation which
contains two undefined functions. Particular choices of the undefined
functions correspond to isentropic flows of an ideal gas, different
forms of the hyperbolic shallow water equations. Complete group classification
of the equation with respect to these functions is performed in \cite{bk:KaptsovMeleshko2019_2}.

The present paper is devoted to constructing invariant solutions of
one of the models of the two-dimensional shallow water equations obtained
in \cite{bk:MeleshkoSamatova2019}. The analysis of these invariant solutions is reduced
to the study of systems of ordinary differential equations. For solving
these systems we used the sixth-order Runge-Kutta numerical method.

\section{THE STUDIED MODEL}

Equations (\ref{eq:June24.1}) contain the arbitrary function $B(x,y)$
and arbitrary constants $f_{0}$ and $\Omega$. Extending the results
of \cite{bk:MeleshkoSamatova2019} to Eulerian coordinates, one finds that the equivalence
group of equations (\ref{eq:June24.1}) has the basis generators
\[
X_{2}^{e}=\partial_{x},\,\,\,X_{5}^{e}=\partial_{t},\,\,\,X_{6}^{e}=\partial_{B},\,\,\,X_{3}^{e}=-\Omega\partial_{f_{0}}+\partial_{y},
\]
\[
X_{1}^{e}=y\partial_{y}+x\partial_{x}-\Omega\partial_{\Omega}+2B\partial_{B},
\]
\[
X_{4}^{e}=y\partial_{y}+x\partial_{x}-2\Omega\partial_{\Omega}-f_{0}\partial_{f_{0}}+t\partial_{t}.
\]
There is also the involution
\[
t\to-t,\,\,\,\Omega\to-\Omega.
\]

Using the equivalence transformation related with the generator $X_{3}$,
one can assume that $f_{0}=0$. Notice also that applying the scaling
corresponding to the generator $X_{4}$ and involution, the constant
$\Omega$ can be reduced to $\Omega=1$.

The group classification of the two-dimensional shallow water equations
with respect to the bottom $B(x,y)$ in mass Lagrangian coordinates
 led to the following results \cite{bk:MeleshkoSamatova2019}. The kernel of admitted Lie
algebras, which consists of the generators admitted by any function
$B(x,y)$, is defined by the generators
\[
-\psi_{\eta}\partial_{\xi}+\psi_{\xi}\partial_{\eta},\,\,\,\partial_{t},
\]
where $\psi(\xi,\eta)$ is an arbitrary function. Extensions of the
kernel are obtained for specific functions $B(x,y)$: the extensions
are presented in Table \ref{tab:Tab1}, where
\[
X_{0}=-t\partial_{t}+6\xi\partial_{\xi}+x\partial_{x}+y\partial_{y}.
\]

\begin{table}
\centering %
\begin{tabular}{c|c|c|l}
\hline
 & $B(x,y)$  & extension  & conditions \tabularnewline
\hline
1  & $y^{4}\Psi(z)+q_{1}x^{4},\ (z=xy^{-1})$  & $X_{0},$  & $z\Psi^{\prime\prime}-3\Psi^{\prime}\neq0$, \tabularnewline
2  & $q_{3}y^{4}+q_{1}x^{4}$  & $X_{0},$  & $q_{1}\neq0$ \tabularnewline
3  & $q_{3}y^{4}-qy^{2}$  & $X_{0}+4t\frac{q}{\Omega}\partial_{x},\,\,\,\partial_{x},$  & \tabularnewline
4  & $2q^{2}x^{2}+\Omega qxy^{2}+q_{1}x+g(y)$  & $e^{2qt}\partial_{x},$  & $q^{2}+q_{1}^{2}\neq0$ \tabularnewline
5  & $g(y)$  & $\partial_{x},$  & $g^{\prime\prime}-3y^{-1}g^{\prime}\neq const$ \tabularnewline
\hline
\end{tabular}\caption{Group classification of equations (\ref{eq:June24.1}) in mass Lagrangian
coordinates.}
\label{tab:Tab1}
\end{table}

In Eulerian coordinates the generators $X_{0}+4t\frac{q}{\Omega}\partial_{x}$
and $e^{2qt}\partial_{x}$ become, respectively,
\[
-t\partial_{t}+(x+4\frac{q}{\Omega}t)\partial_{x}+y\partial_{y}+2(2h\partial_{h}+(2\frac{q}{\Omega}+u)\partial_{u}+v\partial_{v}),
\]
and
\[
e^{2qt}(\partial_{x}+2q\partial_{u}).
\]

In the present paper we consider the model with the maximum extension
which corresponds to the bottom
\[
B(x,y)=q_{3}y^{4}-qy^{2}.
\]
Using the study of \cite{bk:MeleshkoSamatova2019} and extending these results  to Eulerian coordinates,
one obtains that system of equations (\ref{eq:June24.1}) admits the three-dimensional Lie algebra with the basis generators ${\cal L}_{3}=\{X_{1},X_{2},X_{3}\}$, where
\[
X_{1}=\partial_{t},\,\,\,\,X_{3}=\partial_{x},
\]
\[
X_{2}=-t\partial_{t}+(x+4\frac{q}{\Omega}t)\partial_{x}
+y\partial_{y}+2(2h\partial_{h}+(2\frac{q}{\Omega}+u)\partial_{u}+v\partial_{v}).
\]

\section{GROUP CLASSIFICATION OF ${\cal L}_{3}$}

According to the theory of the group analysis method \cite{bk:Ovsiannikov1978},
all invariant solutions are separated into classes of equivalent solutions.
The equivalence is considered with respect to the admitted Lie group
corresponding to the Lie algebra ${\cal L}_{3}$. For finding representatives
of these classes one can use an optimal system of subalgebras
of the admitted Lie algebra ${\cal L}_{3}$ \cite{bk:Ovsiannikov1978}.
As the goal of the present paper is to obtain invariant solutions which are reduced to
a system of ordinary differential equations, then one needs to construct
an optimal system of two-dimensional subalgebras of the Lie algebra
${\cal L}_{3}$.

The commutator table of ${\cal L}_{3}$ is
\[
\begin{array}{c|ccc}
 & X_{1} & X_{2} & X_{3}\\
\hline X_{1} & 0 & -X_{1}+4\frac{q}{\Omega}X_{3} & 0\\
X_{2} & X_{1}-4\frac{q}{\Omega}X_{3} & 0 & -X_{3}\\
X_{3} & 0 & X_{3} & 0
\end{array}
\]
Using the commutator table, one derives the corresponding set of automorphisms
\[
\begin{array}{cl}
A_{1}: & \,\,\,\bar{x}_{1}=x_{1}-ax_{2},\,\,\,\bar{x}_{3}=x_{3}+4\frac{q}{\Omega}ax_{2}\\
A_{2}: & \,\,\,\bar{x}_{1}=x_{1}e^{a},\,\,\bar{x}_{3}=2\frac{q}{\Omega}x_{1}(e^{-a}-e^{a})+x_{3}e^{-a},\\
A_{3}: & \,\,\,\bar{x}_{3}=x_{3}+x_{2}a,
\end{array}
\]
where $x_{i}$ are coefficients of the generator
\[
X=x_{1}X_{1}+x_{2}X_{2}+x_{3}X_{3}.
\]

It is obvious that the two-dimensional subalgebra which does not contain
$X_{2}$ is the only subalebra $\{X_{1},X_{3}\}$.

If one of the generators of the basis generators of the two-dimensional
subalgebra contains the generator $X_{2}$, then it can be represented
as
\[
X_{2}+x_{1}X_{1}+x_{2}X_{2}.
\]
Using the automorphisms $A_{1}$ and then $A_{3}$ this generator
can be reduced to the generator:
\[
X_{2}.
\]
Hence, the second generator can be choseen as $\alpha X_{1}+\beta X_{3}$.
Composing the commutator $[X_{2},\alpha X_{1}+\beta X_{3}]$, one
derives that if $\alpha=1$, then $\beta=-2\frac{q}{\Omega}$. If
$\alpha=0$, then one can choose $\beta=1$. Thus, one obtains the
subalgebras $\{X_{2},X_{1}-2\frac{q}{\Omega}X_{3}\}$ and $\{X_{2},X_{3}\}$.

Therefore, one needs to study the subalgebras
\[
\{X_{1},X_{3}\},\,\,\,\{X_{2},X_{1}-2\frac{q}{\Omega}X_{3}\},\,\,\,\{X_{2},X_{3}\}.
\]

\section{INVARIANT SOLUTIONS}

\subsection{Subalgebra $\{X_{1},X_{3}\}$}

Invariant solutions with respect to these generators describe stationary
one-dimensional solutions:
\[
h=H(y),\,\,\,u=U(y),\,\,\,v=V(y).
\]
The functions $H$, $U$ and $V$ can be found in quadratures
\begin{equation}
\begin{array}{c}
VH=c_{1},\,\,\,U=\Omega y^{2}/2+c_{2},\\
V^{2}+\frac{1}{4}\Omega^{2}y^{4}+\Omega c_{2}y^{2}+4H-2y^{2}(q_{3}y^{2}-q)=c_{3},
\end{array}\label{eq:June24.2}
\end{equation}
where $c_{1}$, $c_{2}$ and $c_{3}$ are constant.

\begin{figure}[h]
\includegraphics[width=.33\textwidth]
{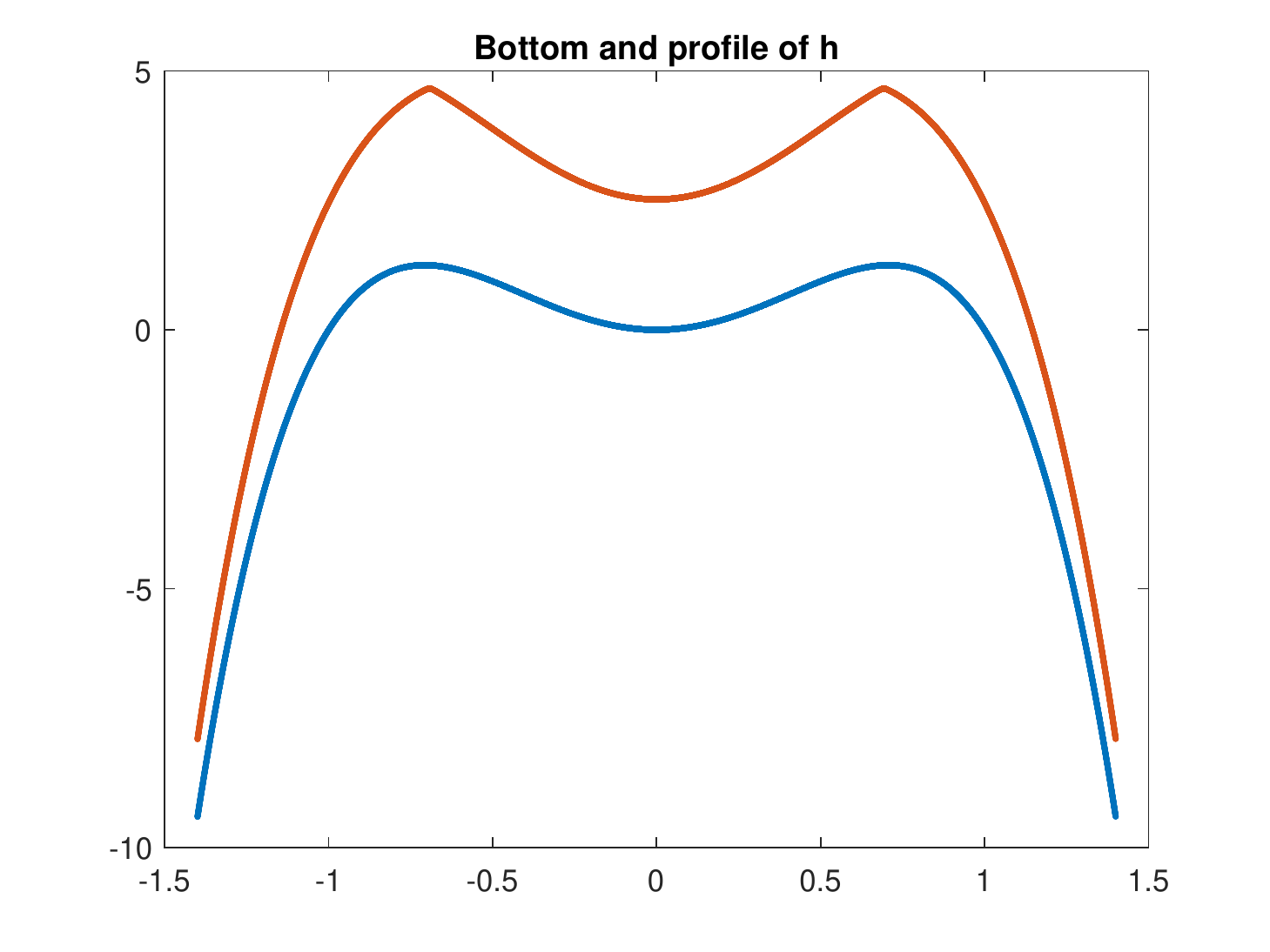}%
\hfill
\includegraphics[width=5.0cm]
{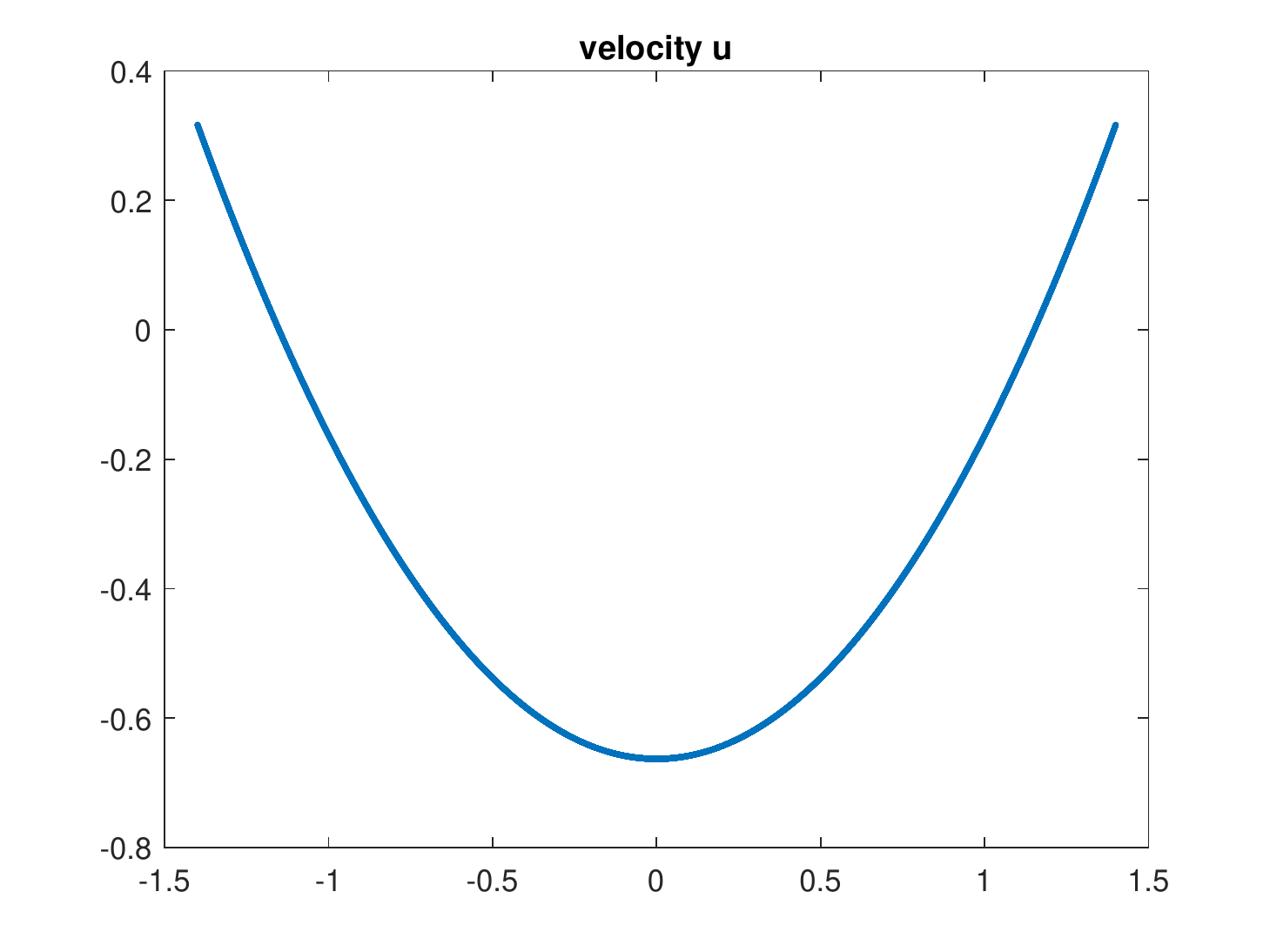}%
\hfill
\includegraphics[width=5.0cm]
{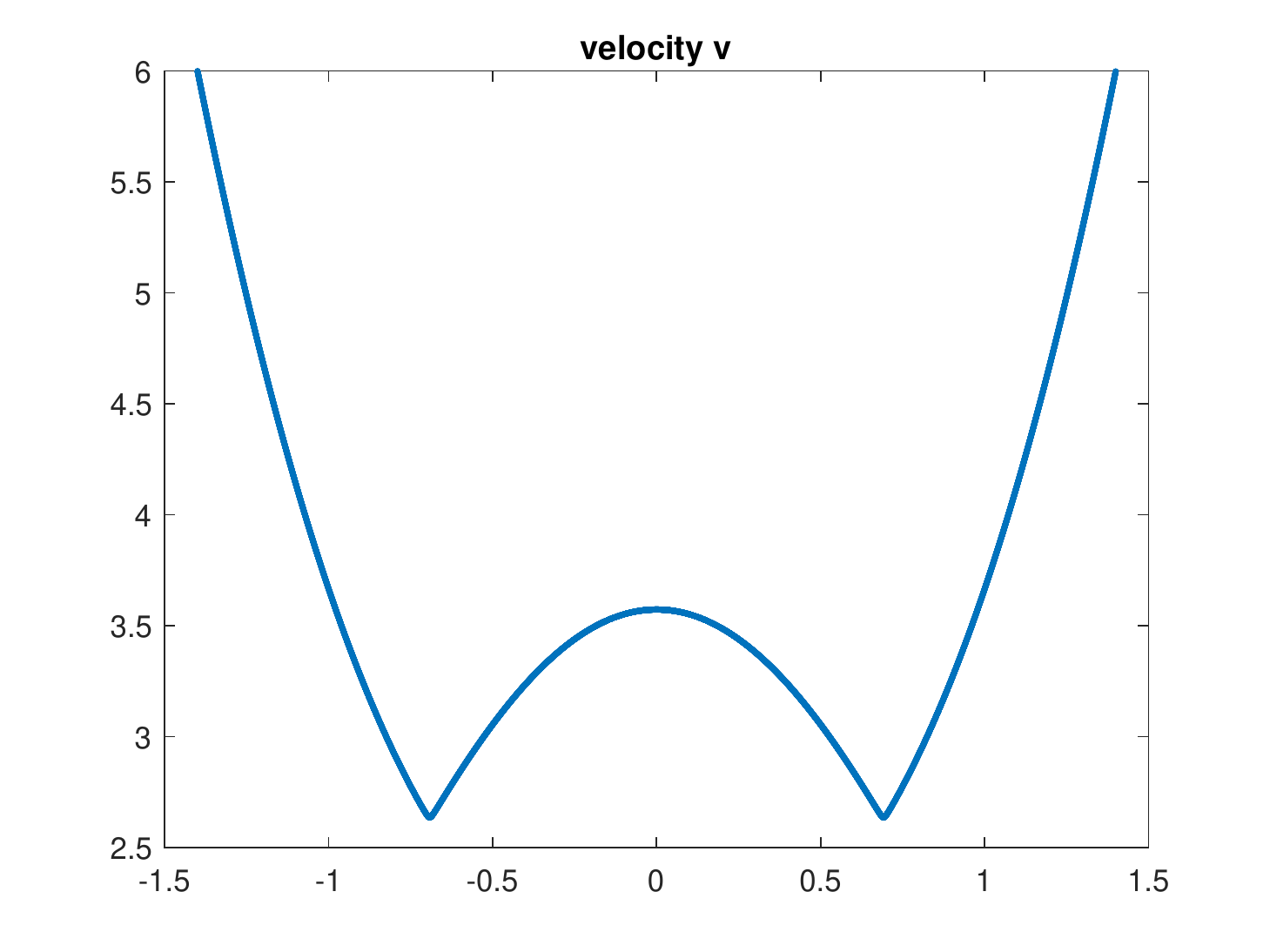}%
\hfill
\small
 \caption{Solution invariant w.r.t. $\{X_1,X_3\}$ with parameters
  $q=5.$ and $q_3=5.$.
\label{fig13}}
 \hfill

 \end{figure}

For finding the function $H(y)$ we used the Runge-Kutta method by solving the ordinary differential equation
\[
H^\prime = H^3y(2c_2\Omega + \Omega^2y^2 + 4q - 8q_3y^2)( 2(c_1^2 - 2H^3)^{-1}.
\]
The calculations were from $a=1.4$ up to $b=-1.4$ with the following parameters  $q_3=5$, $q=5$, $\Omega = 1$, $H(a) = 1.5$, $V(a)= 6$, $U(a)= 0.317$. In calculations we also controlled that the left-hand side of the third equation of (\ref{eq:June24.2}) is constant. Results of these calculations are presented in Fig. \ref{fig13}.
It should be noted that if the velocity  $U(a)$ decreases, for example  $U(a)=0.316$, then the solution is destroyed as  the denominator vanishes at $y=-.7$. If the velocity $U(a)$ increases, then the humps around $y=-.7$ and $y=.7$ become smoother.

\subsection{Subalgebra $\{X_{2},X_{1}-2\frac{q}{\Omega}X_{3}\}$}

\begin{figure}[h]
\includegraphics[width=.3\textwidth]
{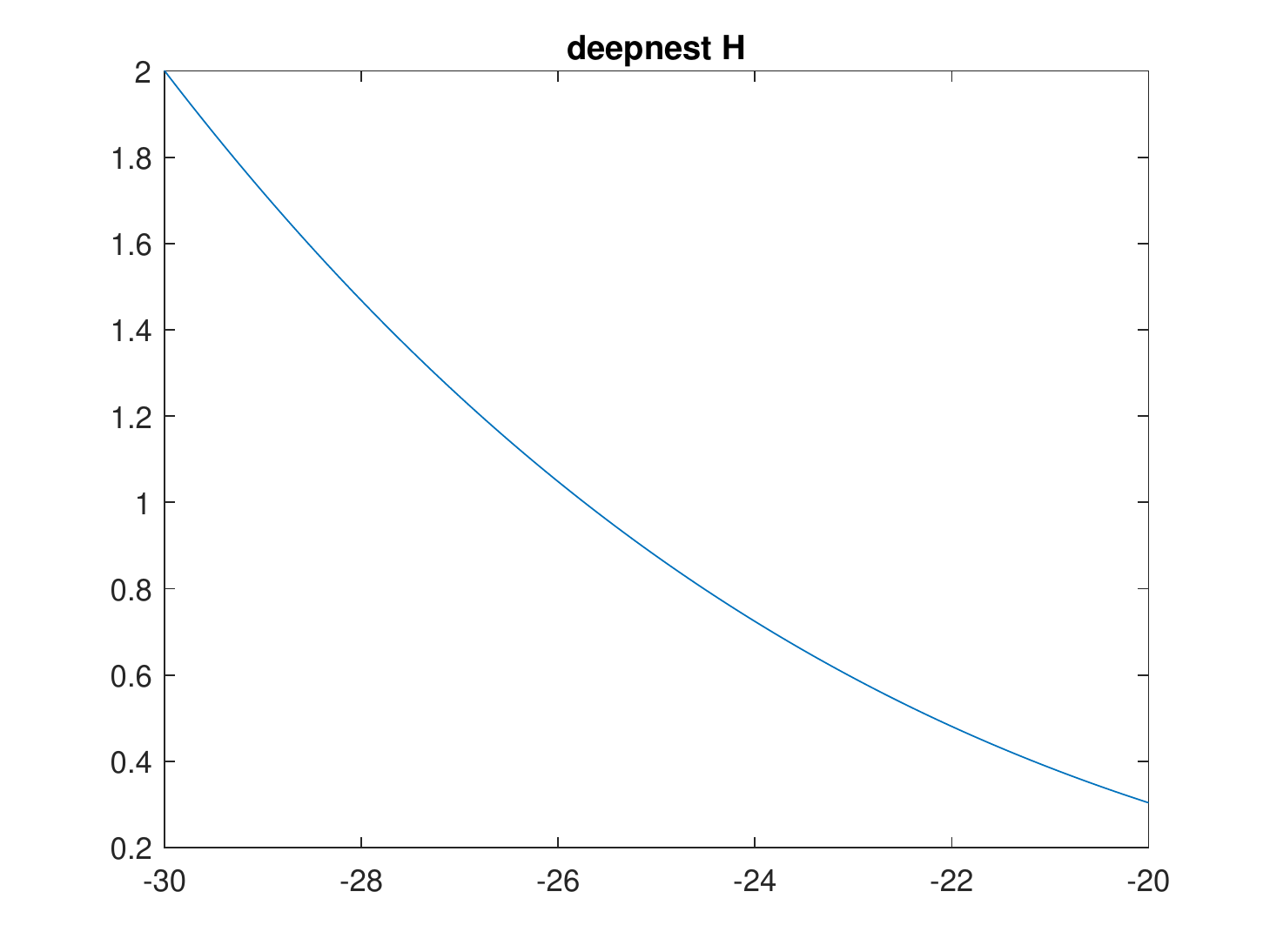}%
\hfill
\includegraphics[width=.3\textwidth]
{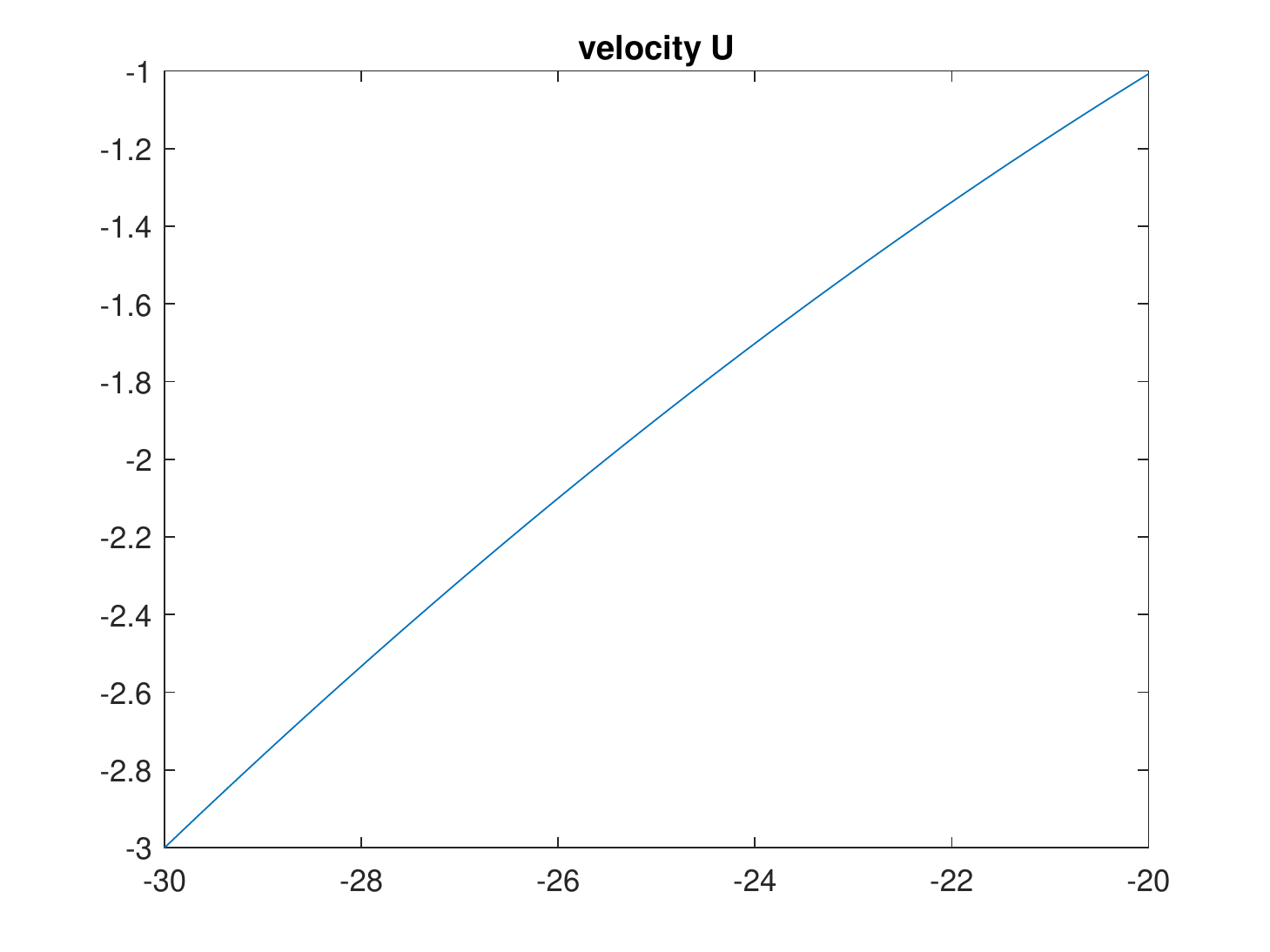}%
\hfill
\includegraphics[width=.3\textwidth]
{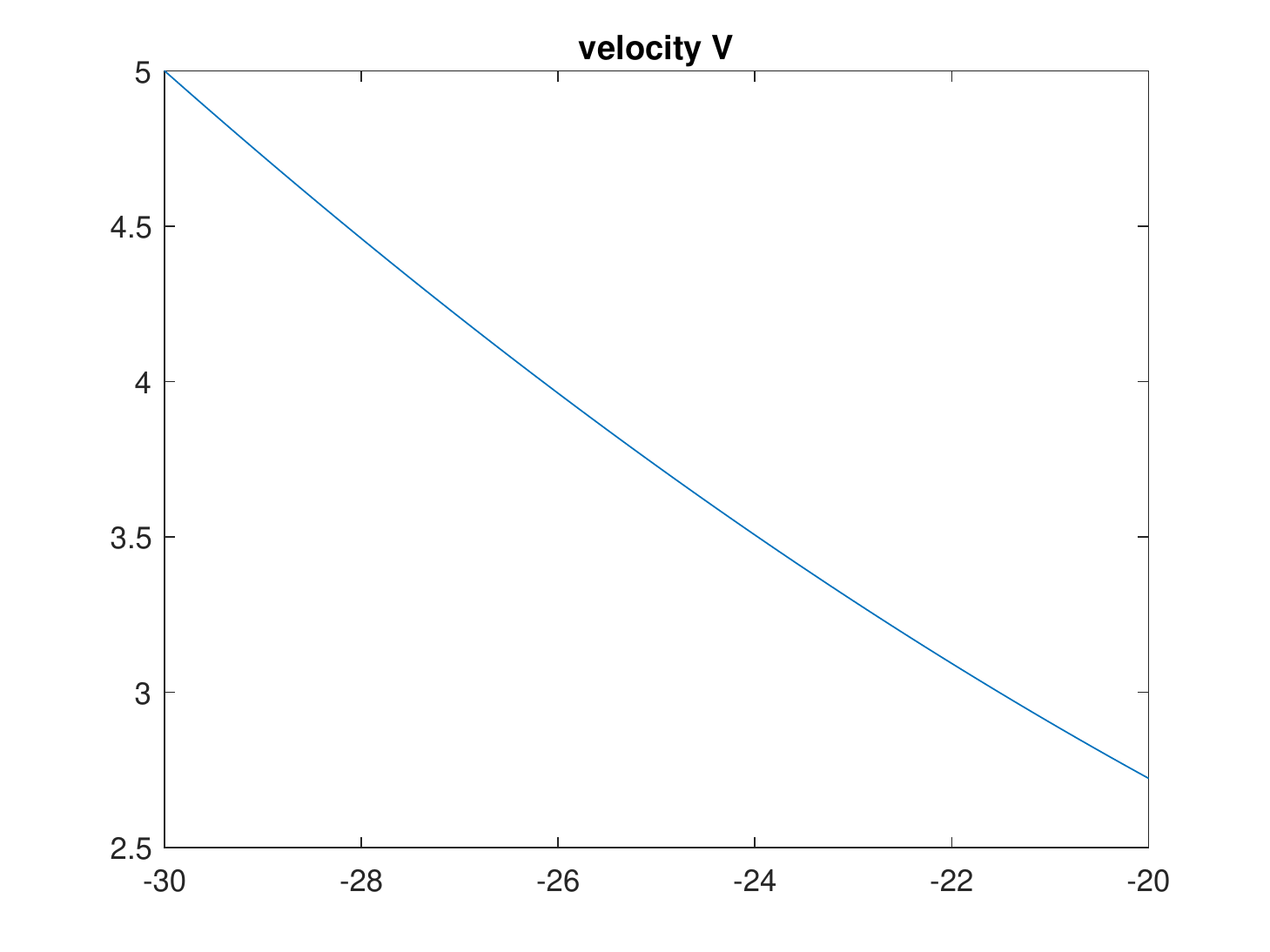}%
\hfill
\small
 \caption{Solution invariant w.r.t. $\{X_{2},X_{1}-2\frac{q}{\Omega}X_{3}\}$ with the parameter $q_3=5$.
\label{fig46}}
 \hfill

 \end{figure}

Invariants of this subalgebra are
\[
(x+2\frac{q}{\Omega}t)y^{-1},\,\,\,hy^{-4},\,\,\,(u+2\frac{q}{\Omega})y^{-2},\,\,\,vy^{-2}.
\]
Hence, a representation of an invariant solution is
\[
h=y^{4}H(z),\,\,\,u=-2\frac{q}{\Omega}+U(z),\,\,\,v=y^{2}V(z),
\]
where $z=(x+2\frac{q}{\Omega}t)y^{-1}$. The reduced system of equations
becomes
\begin{equation}{c}
\label{eq:31July.1}
H^\prime = \frac{N_h}{D_h},\ U^\prime = \frac{N_u}{D_u},\ V^\prime = \frac{N_v}{D_u},
\end{equation}
where
\[
N_h =  8 H ^2 z  + H  (\Omega  U  z  + \Omega  V  - 4 q_3 z  + 4 U  V  - 4 V ^2 z ),
\]
\[
D_h = 2 H  (z ^2 + 1) - (U - V z)^2,
\]
\[
\begin{array}{c}
N_u  =   - 16 H ^2 z  + 2 H  ( - \Omega  U  z  + \Omega  V  z ^2 + 4 q_3 z
- 2 U  V  z ^2 - 6 U  V  + 4 V ^2 z )
\\
+ V  ( - \Omega  U ^2 + 2 \Omega  U  V  z  - \Omega  V ^2 z ^2 + 2 U ^3 - 4 U ^2 V  z  + 2 U  V ^2 z ^2),
\end{array}
\]
\[
D_u = (U  - V  z ) D_h,
\]
\[
\begin{array}{c}
N_v  =   - 16 H ^2 + 2 H  ( - \Omega  U  + \Omega  V  z  + 4 q_3 + 4 U ^2 - 4 U  V  z
- 2 V ^2 z ^2 - 2 V ^2)
\\
+ \Omega  U ^3 - 2 \Omega  U ^2 V  z  + \Omega  U  V ^2 z ^2 - 4 q_3 U ^2 + 8 q_3 U  V  z
- 4 q_3 V ^2 z ^2
\\
+ 2 U ^2 V ^2 - 4 U  V ^3 z  + 2 V ^4 z ^2,
\end{array}
\]
Notice that the right-hand sides of equations (\ref{eq:31July.1}) do not depend on $q$.
The solution has some similarities with a travelling wave type solutions.
The results of calculations with $U(-30)=-3$, $V(-30)=5$,
$H(-30)=2$ are given in Fig. \ref{fig46}.

\subsection{Subalgebra $\{X_{2},X_{3}\}$}

\begin{figure}[h]
\includegraphics[width=.3\textwidth]
{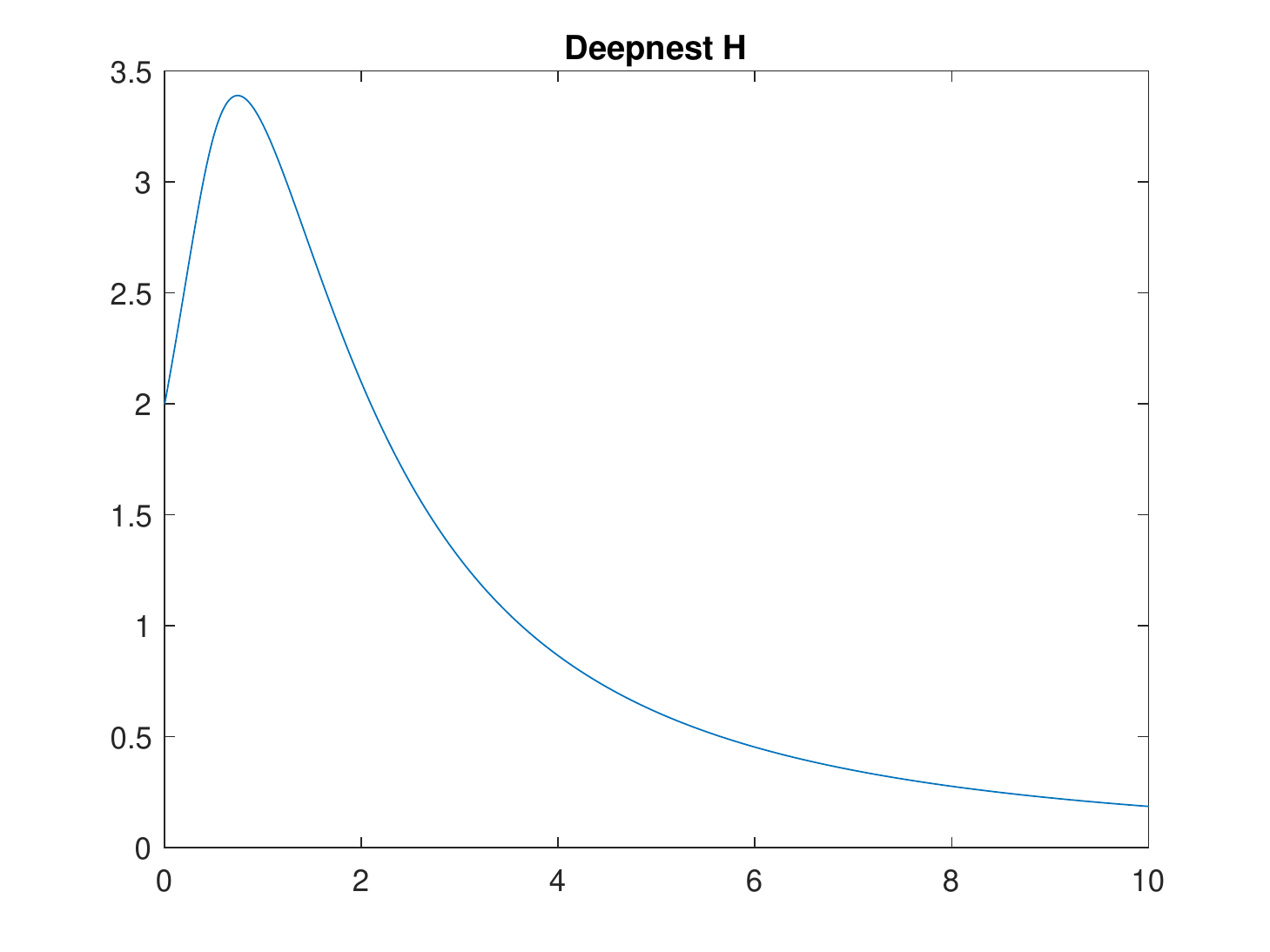}%
\hfill
\includegraphics[width=.3\textwidth]
{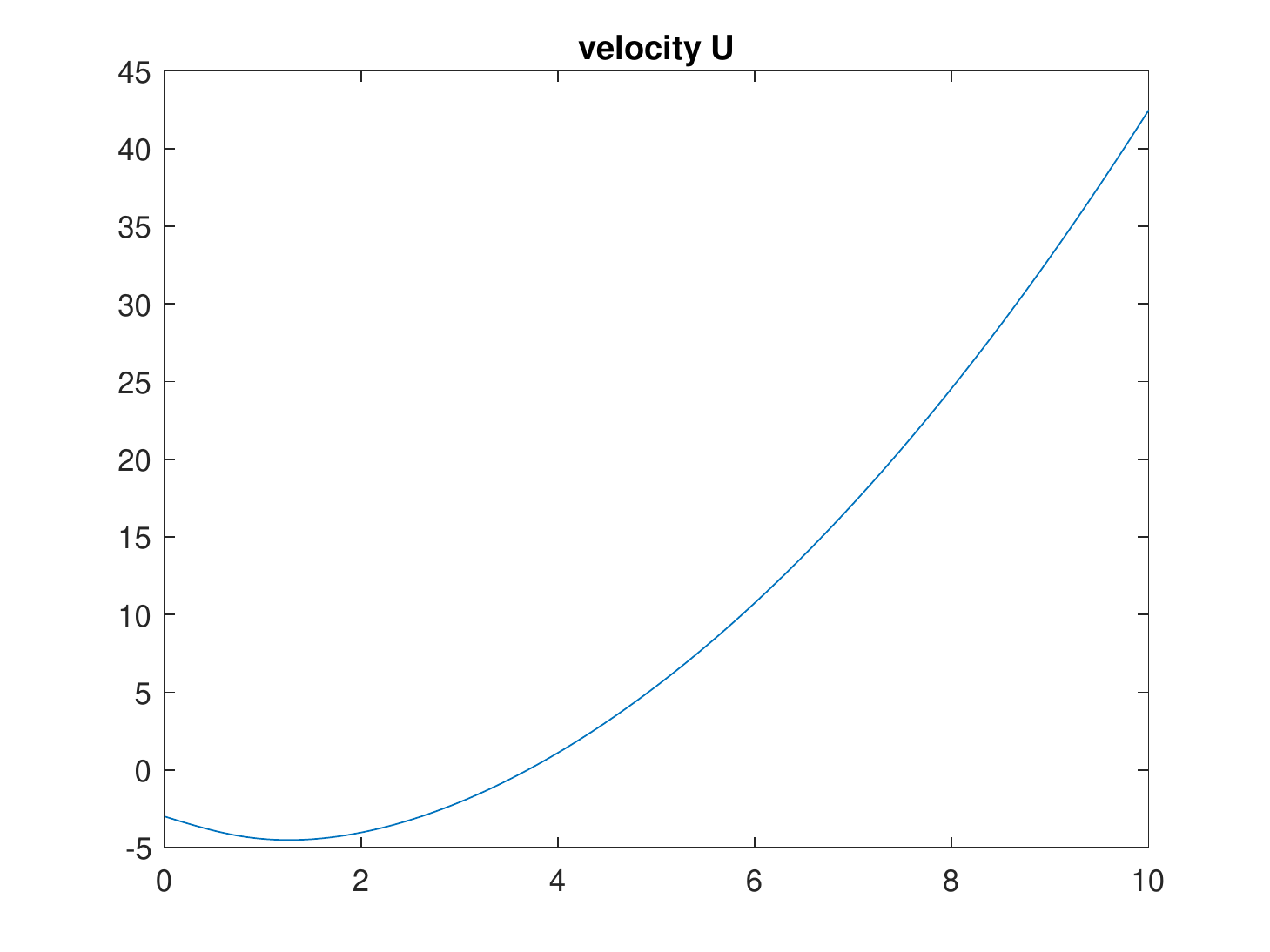}%
\hfill
\includegraphics[width=.3\textwidth]
{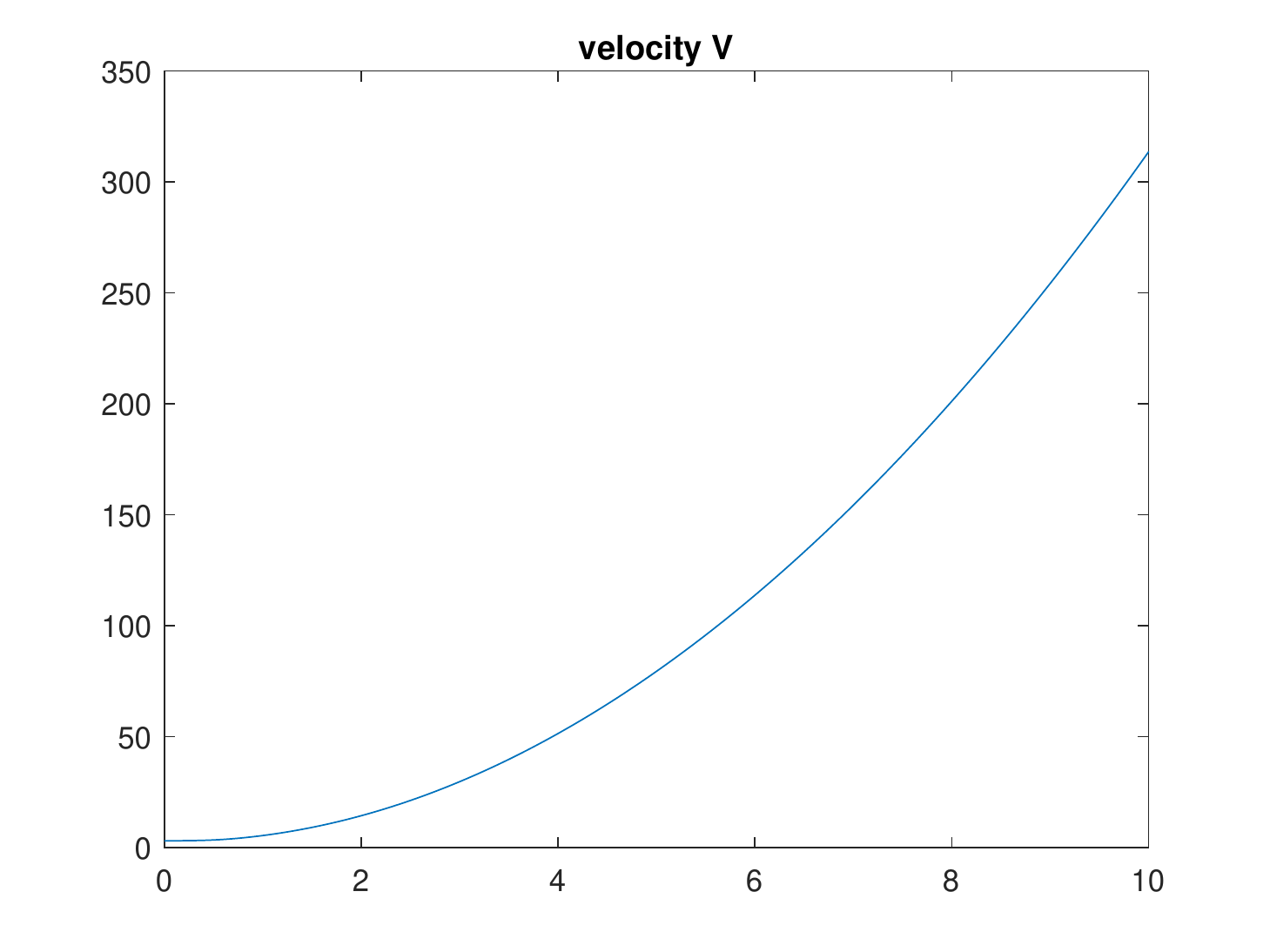}%
\hfill
\small
 \caption{Solution invariant w.r.t. $\{X_2,X_3\}$ with the parameter
  $q_3=1$.
\label{fig79}}
 \hfill

 \end{figure}

Invariants of this subalgebra are
\[
yt,\,\,\,ht^{4},\,\,\,(u+2\frac{q}{\Omega})t^{2},\,\,\,vt^{2}.
\]
Hence, representation of an invariant solution is
\[
h=t^{4}H(z),\,\,\,u=-2\frac{q}{\Omega}+t^{-2}U(z),\,\,\,v=t^{-2}V(z),
\]
where $z=yt$. The reduced system of equations
becomes
\begin{equation}
\label{eq:jul30.2}
\begin{array}{c}
H^\prime = (2 H - (V+z)^2)^{-1}( 4 H z  (q_3 z ^2 - 1) - H U z  - 2 H V ),
\\
U^\prime = (V  + z )^{-1}(2 U + V  z ),
\\
V ^\prime = (2 H - (V+z)^2)^{-1}(8 H + U V  z  + U z ^2 - 2 V ^2 - 2 V  z  ( 2 q_3 z ^2 + 1)- 4 q_3 z ^4).
\end{array}
\end{equation}
 It should be noted that the right-hand sides of equations (\ref{eq:jul30.2}) also do not depend on $q$.
The results of calculations with $U(-30)=-3$, $V(-30)=5$,
$H(-30)=2$ are given in Fig. \ref{fig79}.

\section{CONCLUSION}

The main result of the present paper is  the complete set of invariant solutions of system (\ref{eq:June24.1}) for which a reduced system is a system of ordinary differential equations. There are three such classes of solutions. Analysis of these invariant solutions is performed by the Runge-Kutta numerical method. It was noted that the parameter $q$ does not have an influence on deriving the functions $H$, $U$ and $V$ in cases of $\{X_{2},X_{1}-2\frac{q}{\Omega}X_{3}\}$ and $\{X_{2},X_{3}\}$, whereas in the case $\{X_{1},X_{3}\}$ it has substantial influence.
In all cases the numerical solutions are destroyed when the corresponding denominator was vanishing.

\section*{ACKNOWLEDGEMENTS}

The research was supported by Russian Science Foundation Grant No
18-11-00238 `Hy\-dro\-dynamics-type equations: symmetries, conservation
laws, invariant difference schemes'.


\end{document}